\documentclass[letterpaper,titlepage,11pt]{article}
\usepackage{hyperref}
\usepackage{amssymb,amsmath,amsfonts}
\usepackage{epsfig}

\def\clock{{\count0=\time
           \divide\count0 60
           \ifnum\count0<10 0\fi\the\count0
           \multiply\count0 -60 \advance\count0 \time
           :\ifnum\count0<10 0\fi \the\count0
         }}
\newcommand{\timestamp}{{\small\vbox{\hbox{\tt\jobname.tex}
\hbox{\the\day/\the\month/\the\year, \clock}}}}

\newcommand{\be}{\begin{equation}} \newcommand{\ee}{\end{equation}}
\newcommand{\bea}{\begin{eqnarray}} \newcommand{\eea}{\end{eqnarray}}

\newcommand{\CO}{\mathcal{O}}

\newcommand{\CT}{\mathcal{T}}

\newcommand{\CM}{\mathcal{M}}

\newcommand{\id}{\hbox{1\kern-.27em l}}
\newcommand{\sid}{\hbox{\scriptsize1\kern-.27em l}}

\newcommand{\we}{\kern-.1em\wedge\kern-.1em}
\newcommand{\scal}{\kern-.13em\cdot\kern-.13em}

\newcommand{\II}{I\kern-.09em I}

\newcommand{\R}{\mathbb{R}}

\newcommand{\spa}{\ , \ \ }

\newcommand{\mt}{\mathfrak{t}}
\newcommand{\ms}{\mathfrak{s}}

\newcommand{\beastar}{\begin{eqnarray*}}
\newcommand{\eeastar}{\end{eqnarray*}}
\newcommand{\rom}[1]{\mathrm{#1}}
\numberwithin{equation}{section}

\begin{document}

\begin{titlepage}

\rightline{\vbox{\small\hbox{\tt hep-th/0503020} }} \vskip 3cm

\centerline{\Large \bf Phases of Kaluza-Klein Black Holes:} \vskip
0.2cm \centerline{\Large \bf A Brief Review}

\vskip 1.6cm \centerline{\bf Troels Harmark and Niels A. Obers}
\vskip 0.5cm
\begin{center}
\sl The Niels Bohr Institute \\
\sl Blegdamsvej 17, 2100 Copenhagen \O , Denmark
\end{center}

\vskip 0.5cm

\centerline{\small\tt harmark@nbi.dk, obers@nbi.dk}

\vskip 1.6cm

\centerline{\bf Abstract} \vskip 0.2cm \noindent We review the
latest progress in understanding the phase structure of static and
neutral Kaluza-Klein black holes, i.e. static and neutral
solutions of pure gravity with an event horizon that asymptote to
a $d$-dimensional Minkowski-space times a circle. We start by reviewing the
$(\mu,n)$ phase diagram and the split-up of the phase structure
into solutions with an internal $SO(d-1)$ symmetry and solutions
with Kaluza-Klein bubbles. We then discuss the uniform black string,
non-uniform black string and localized black hole phases, and how
those three phases are connected, involving issues such as
classical instability and horizon-topology changing transitions.
Finally, we review the bubble-black hole sequences, their
place in the phase structure and interesting aspects such as the
continuously infinite non-uniqueness of solutions for a given mass
and relative tension.


\end{titlepage}

\pagestyle{empty} \small
\normalsize

\pagestyle{plain} \setcounter{page}{1}

\section{Introduction}

In this brief review we go through recent progress
in the understanding of static and neutral Kaluza-Klein black holes.%
\footnote{This review is an extended and updated version of the proceedings
published in Class.Quant.Grav.21:S1509-S1516,2004, for a
talk given by T. Harmark at
the EC-RTN workshop ``The quantum structure of spacetime and the
geometric nature of fundamental interactions'' held September 2003
in Copenhagen.}
 A $d+1$ dimensional static and neutral
Kaluza-Klein black hole is defined here as a pure gravity solution
that asymptotes to $d$-dimensional Minkowski-space times a circle
$\CM^d \times S^1$ at infinity, with at least one event horizon.%
\footnote{In other words, we consider solutions of the vacuum
Einstein equations for which we have at least one event horizon
present, and such that the asymptotics of the solution is
$(d+1)$-dimensional Kaluza-Klein space $\CM^d \times S^1$.} Since
we consider only static and neutral solutions they do not have
charges or angular momenta.

Static and neutral Kaluza-Klein black holes are interesting to
study for a variety of reasons:
\begin{itemize}
\item Four-dimensional static and neutral black holes have a very
simple phase-structure. For a given mass, there is only one phase
available: The Schwarzschild black hole. The phase structure of
Kaluza-Klein black holes, on the other hand, is very rich, as we
review below. Indeed, for some choice of masses we have a
continuous amount of different Kaluza-Klein black holes with that
mass.

\item The phase structure of Kaluza-Klein black holes contain
examples of phase transitions, which are therefore purely gravitational
phase transitions between different solutions with event horizons.
Specifically, one has examples of phase transitions involving
topology change of the event horizon.
One of the most interesting examples is the decay of the uniform black string to
the localized black hole solution.

\item If we have large extra dimensions in Nature the Kaluza-Klein black
hole solutions, or generalizations thereof, will become relevant
for experiments involving microscopic black holes and observations
of macroscopic black holes.

\end{itemize}

As will be discussed in the Conclusion, Kaluza-Klein black holes
are also intimately related to non- and near-extremal branes of
String/M-Theory with a circle in the transverse space. Via the
gauge/gravity duality this means that the study of Kaluza-Klein
black holes is also relevant for the thermal phase structure
of the non-gravitational theories that are dual to near-extremal
branes with a circle in the transverse space. This development
is very briefly reviewed in the proceedings \cite{Harmark:2005xx}.

See Ref.~\cite{Kol:2004ww} for another  recent review of
Kaluza-Klein black holes which contains complimentary motivations and
discussions.


\section{Physical parameters and definition of phase diagram}
\label{sec:phase}

In this section we show how to measure the mass $\mu$ and relative tension $n$
of a Kaluza-Klein black hole, so that each phase can
be plotted in the $(\mu,n)$ phase diagram. We review the main
features of the split-up of this phase diagram into two regions.
Finally, we discuss some general results for the thermodynamics of Kaluza-Klein
black holes.

\subsection*{Measuring mass and tension and defining phase diagram}

For any space-time which is asymptotically $\CM^d \times S^1$ we
can measure the mass $M$ and the tension $\CT$. These quantities
can then conveniently be used to display the various phases of
Kaluza-Klein black holes, as we review below.

Define the Cartesian coordinates for $d$-dimensional Minkowski
space $\CM^d$ as $t,x^1,...,x^{d-1}$ and the radius $r =
\sqrt{\sum_i (x^i)^2 }$. In addition we have the coordinate $z$
for the $S^1$, of period $L$. Thus, the total space-time dimension
is $D=d+1$. Now, for any localized static object we have the
asymptotics \cite{Harmark:2003dg}
\begin{equation}
\label{gttzz} g_{tt} \simeq - 1 + \frac{c_t}{r^{d-3}} \spa g_{zz} \simeq 1 +
\frac{c_z}{r^{d-3}} \ ,
\end{equation}
for $r \rightarrow \infty$. The mass $M$ and tension $\CT$ are then
given by \cite{Harmark:2003dg,Kol:2003if}
\begin{equation}
\label{MT} M = \frac{\Omega_{d-2} L}{16 \pi G_{\rm N}} \left[
(d-2) c_t - c_z \right] \spa \CT = \frac{\Omega_{d-2} }{16 \pi
G_{\rm N}} \left[ c_t - (d-2) c_z \right] \ .
\end{equation}
The tension has previously been defined in
\cite{Traschen:2001pb,Townsend:2001rg}. The tension in \eqref{MT}
can also be obtained from the general formula for tension in terms
of the extrinsic curvature \cite{Harmark:2004ch}, analogous to the
Hawking-Horowitz mass formula \cite{Hawking:1996fd}. The mass and
tension formulas have been generalized to non-extremal and
near-extremal branes in \cite{Harmark:2004ws}.

We now define the {\sl relative tension} (also called
 the {\sl relative binding energy}) as
\cite{Harmark:2003dg}
\begin{equation}
\label{then} n = \frac{\CT L}{M} = \frac{c_t - (d-2) c_z}{(d-2)
c_t - c_z } \ .
\end{equation}
This measures how large the tension (or binding energy) is
relative to the mass. This is a useful quantity since it is
dimensionless and since it is bounded as \cite{Harmark:2003dg}
\begin{equation}
\label{nbound} 0 \leq n \leq d-2 \ .
\end{equation}
Here the upper bound is due to the Strong Energy Condition while
the lower bound was found in
\cite{Traschen:2003jm,Shiromizu:2003gc}. The upper bound can also
be  physically understood in a more direct way. If we consider a
test particle at infinity it is easy to see that the gravitational
force on the particle is attractive for $n < d-2$ but repulsive
for $n > d-2$.

It is also useful to define a rescaled dimensionless quantity from
the mass as
\begin{equation}
\label{themu} \mu = \frac{16\pi G_{\rm N}}{L^{d-2}} M \ .
\end{equation}
We can now formulate the program set forth in
\cite{Harmark:2003dg,Harmark:2003eg}: To plot all phases of
Kaluza-Klein black holes in a $(\mu,n)$ diagram.

\subsection*{The split-up of the phase diagram}

According to the present knowledge of phases of static and neutral
Kaluza-Klein black holes, the $(\mu,n)$ phase diagram appears to
be divided into two separate regions \cite{Elvang:2004iz}:
\begin{itemize}
\item The region $0 \leq n \leq 1/(d-2)$ contains solutions
without Kaluza-Klein bubbles, and the solutions have a local
$SO(d-1)$ symmetry. We review what is know about solutions in this
part of the phase diagram in Section \ref{sec:bhc}. The general
properties of this class of solutions, which we in the following
call black holes and strings on cylinders (since $\CM^d \times
S^1$ is the cylinder $\R^{d-1} \times S^1$ for a fixed time), are
the subject of \cite{Harmark:2003dg,Harmark:2003eg}. Due to the
$SO(d-1)$ symmetry there are only two types of event horizon
topologies: The event horizon topology is $S^{d-1}$ ($S^{d-2}
\times S^1$) for the black hole (string) on a cylinder.
\item The region $1/(d-2) < n \leq d-2$ contains solutions with
Kaluza-Klein bubbles. We review this class of solutions in Section
\ref{sec:bub}. This part of the phase diagram is the subject of
\cite{Elvang:2004iz}. It turns out that this part of the phase diagram
is much more densely populated with solutions than the lower part.
\end{itemize}

\subsection*{Thermodynamics, first law and the Smarr formula}

For a neutral Kaluza-Klein black hole with a single connected
horizon, we can find the temperature $T$ and entropy $S$ directly
from the metric. Together with the mass $M$ and relative tension
$n$, these quantities obey the Smarr formula
\cite{Harmark:2003dg,Kol:2003if} $(d-1) TS = (d-2-n)M$ and the
first law of thermodynamics \cite{Kol:2003if,Harmark:2003yz} $dM =
TdS + n M dL/L$. It is useful to define the rescaled temperature
$\mt$ and entropy $\ms$ by
\begin{equation}
\label{tsneut} \mt = L T \spa \ms = \frac{16 \pi G_{\rm
N}}{L^{d-1}} S \ .
\end{equation}
In terms of these quantities, the Smarr formula for Kaluza-Klein
black holes and the first law of thermodynamics take the form
\begin{equation}
(d-1) \mt \ms = (d-2-n) \mu  \spa \delta \mu = \mt \, \delta \ms \
.
\end{equation}
Combining the Smarr formula and the first law, we get
the useful relation
\begin{equation}
\label{neutfirst2} \frac{\delta \log \ms}{\delta \log \mu} =
\frac{d-1}{d-2-n} \ ,
\end{equation}
so that, given a curve $n(\mu)$ in the phase diagram, the entire
thermodynamics can be obtained.

\begin{figure}[ht]
\centerline{\epsfig{file=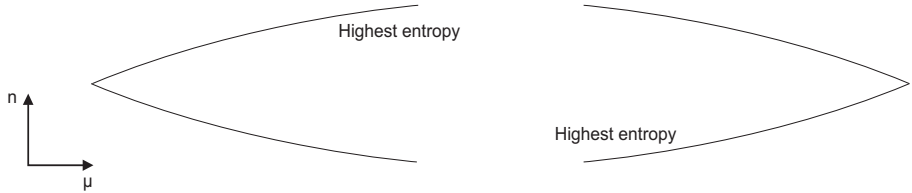,width=12
cm,height=2.5cm}} \caption{Illustration of the Intersection Rule.}
 \label{fig_intrule} \end{figure}

From \eqref{neutfirst2} it is also possible to derive the {\sl
Intersection Rule} of \cite{Harmark:2003dg}: For two branches that
intersect in the same solution, we have the property that for
masses below the intersection point the branch with the lower
relative tension has the highest entropy, whereas for masses above
the intersection point it is the branch with the higher relative
tension that has the highest entropy. We have illustrated the
Intersection Rule in Figure \ref{fig_intrule}.

As seen in \cite{Elvang:2004iz}, there are also neutral
Kaluza-Klein black hole solution with more than one connected
event horizon. The generalization of the Smarr formula and first
law were found in \cite{Elvang:2004iz} for the specific class of
solutions considered there.

\section{Black holes and strings on cylinders}
\label{sec:bhc}

In this section we review the neutral and static black objects without
Kaluza-Klein bubbles. These turn out to have local $SO(d-1)$ symmetry,
which means that the solutions fall into two classes, black holes
with event horizon topology $S^{d-1}$ and black strings with event
horizon topology $  S^{d-2}\times S^1$. We first discuss the ansatz
that follows from the local $SO(d-1)$ symmetry.
We then review in turn the uniform black string,
non-uniform black string and localized black hole phases. These phases
are drawn in the $(\mu,n)$ phase diagram for the five- and six-dimensional
cases. Finally we consider various topics, including copies of solutions,
the endpoint of the decay of the uniform black string and an observation
related to the large $d$ behavior.

\subsection*{The ansatz}

As mentioned above, all solutions with $0 \leq n \leq 1/(d-2)$
have, to our present knowledge, a local $SO(d-1)$ symmetry. Using
this symmetry it has been shown
\cite{Wiseman:2002ti,Harmark:2003eg} that the metric of these
solutions can be written in the form
\begin{equation}
\label{ansatz} ds^2 = - f dt^2 + \frac{L^2}{(2\pi)^2} \left[
\frac{A}{f} dR^2 + \frac{A}{K^{d-2}} dv^2 + K R^2 d\Omega_{d-2}^2
\right] \spa f = 1 - \frac{R_0^{d-3}}{R^{d-3}} \ ,
\end{equation}
where $R_0$ is a dimensionless parameter, $R$ and $v$ are
dimensionless coordinates and the metric is determined by the two
functions $A=A(R,v)$ and $K=K(R,v)$. The form \eqref{ansatz} was
originally proposed in Ref.~\cite{Harmark:2002tr,Harmark:2003fz} as an ansatz for
the metric of black holes on cylinders.

The properties of the ansatz \eqref{ansatz} were extensively
considered in \cite{Harmark:2002tr}. It was found that the
function $A=A(R,v)$ can be written explicitly in terms of the
function $K(R,v)$ thus reducing the number of free unknown
functions to one. The functions $A(R,v)$ and $K(R,v)$ are periodic
in $v$ with the period $2\pi$. Note that $R = R_0$ is the location
of the horizon in \eqref{ansatz}.

As already stated, all phases without Kaluza-Klein bubbles have,
to our present knowledge, $0 \leq n \leq 1/(d-2)$ and can be
described by the ansatz \eqref{ansatz} due to their local
$SO(d-1)$ symmetry. In the following we review the three known
phases and describe their properties.

\subsection*{Uniform string branch}

The uniform string branch consists of neutral black strings which
are translationally invariant along the circle direction. The
metric of a uniform string in $d+1$ dimensions is
\cite{Tangherlini:1963}%
\footnote{The metric \eqref{unmet} corresponds to
$A(R,v)=K(R,v)=1$ in the ansatz \eqref{ansatz}.}
\begin{equation}
\label{unmet} ds^2 = - \left( 1 - \frac{r_0^{d-3}}{r^{d-3}}
\right) dt^2 + \left( 1 - \frac{r_0^{d-3}}{r^{d-3}} \right)^{-1}
dr^2 + r^2 d\Omega_{d-2}^2 + dz^2 \ .
\end{equation}
In terms of the relative binding energy $n$ defined above, we note
that $n=1/(d-2)$ for all of the uniform string branch, as can be
seen from the metric \eqref{unmet} using \eqref{gttzz} and
\eqref{then}. Thus, the uniform string branch is a horizontal line
in the $(\mu,n)$ diagram. The rescaled entropy of the uniform black string is
given by
\begin{equation}
\label{su}
\ms_{\rm u} (\mu) = C_1^{(d-1)} \mu^{\frac{d-2}{d-3}} \ ,
\end{equation}
where the constant $C_1^{(q)}$ is defined as
\begin{equation}
\label{Cdef}
C_1^{(q)} \equiv 4\pi (\Omega_{q-1})^{-\frac{1}{q-2}} (q-1)^{-\frac{q-1}{q-2}} \ .
\end{equation}
The horizon topology of the uniform black string
is clearly $S^1 \times S^{d-2}$,
where the $S^1$ is along the circle-direction.

The most important physical feature of the neutral uniform black
string branch is that it can be classically unstable. Gregory and
Laflamme showed in \cite{Gregory:1993vy,Gregory:1994bj} that a
neutral uniform black string is classically unstable for $\mu <
\mu_{\rm GL}$, i.e. for sufficiently small masses. See Table
\ref{tabnonuni} for numerical values of the critical mass
$\mu_{\rm GL}$ for $d \leq 14$. For $\mu
> \mu_{\rm GL}$ the string is believed to be classically stable. We
comment on the endpoint of the classical instability of the
uniform black string below.

\subsection*{Non-uniform string branch}

As realized in \cite{Gregory:1988nb,Gubser:2001ac} the classical
instability of the uniform black string for $\mu < \mu_{\rm GL}$
points to the existence of a marginal mode at $\mu =\mu_{\rm GL}$
which again suggests the existence of a new branch of solutions.
Since this new branch of solutions should be continuously
connected to the uniform black string it should have the same
horizon topology, at least when the deformation away from the
uniform black string is sufficiently small. Moreover, the solution
should be non-uniformly distributed in the circle-direction $z$
since the marginal mode has a dependence on this direction.

This new branch, here called the non-uniform string branch, has
been found numerically
\cite{Gubser:2001ac,Wiseman:2002zc,Sorkin:2004qq}. Essential
features are:
\begin{itemize}
\item The horizon topology is $S^1 \times S^{d-2}$ with the $S^1$
being supported by the circle of the Kaluza-Klein space-time
$\CM^d \times S^1$. Therefore the solutions describe black
strings.
\item The solutions are non-uniformly distributed along
$z$.
\item The non-uniform black strings have a local $SO(d-1)$
symmetry and can therefore be written in terms of the ansatz
\eqref{ansatz} \cite{Wiseman:2002ti,Harmark:2003eg}.
\item The
non-uniform string branch meets the uniform string branch at
$\mu=\mu_{\rm GL}$, i.e. with $n=1/(d-2)$.
\item The branch has $n
< 1/(d-2)$.
\end{itemize}
If we consider the non-uniform black string branch for
$|\mu-\mu_{\rm GL}| \ll 1$, we have
\begin{equation}
\label{nofmu}
n(\mu) = \frac{1}{d-2} - \gamma ( \mu - \mu_{\rm GL}) + \CO ( (
\mu - \mu_{\rm GL})^2 ) \ .
\end{equation}
Here $\gamma$ is a number representing the slope of the curve
corresponding to the non-uniform string branch near $\mu=\mu_{\rm GL}$.
In Table \ref{tabnonuni} we have listed numerical data for $\mu_{\rm GL}$ and $\gamma$
for $4 \leq d \leq 14$. These data were computed in \cite{Gregory:1993vy,Gregory:1994bj,Gubser:2001ac,Wiseman:2002zc,Sorkin:2004qq}.\footnote{$\gamma$ in Table \ref{tabnonuni} is found
in terms of $\eta_1$ and $\sigma_2$ given in Figure 2 of \cite{Sorkin:2004qq}
by the formula
\[
\gamma = - \frac{2(d-1)(d-3)^2 }{(d-2)^2}
\frac{\sigma_2}{(\eta_1)^2} \frac{1}{\mu_{\rm GL}} \ .
\]
$\eta_1$ and $\sigma_2$ are also found in \cite{Gubser:2001ac,Wiseman:2002zc}
for $d=4,5$.}

\begin{table}[ht]
\begin{center}
\begin{tabular}{|c||c|c|c|c|c|c|c|c|c|c|c|}
\hline $d$ & $4$ & $5$ & $6$ & $7$ & $8$ & $9$ & $10$ & $11$ &
$12$ & $13$ & $14$
\\ \hline
$\mu_{\rm GL}$ & $3.52$ & $2.31$ & $1.74$ & $1.19$ & $0.79$ & $0.55$ & $0.37$ & $0.26$ & $0.18$ & $0.12$ & $0.08$ \\
\hline
$\gamma$  & $0.14$ & $0.17$ & $0.21$ & $0.31$ & $0.47$ & $0.74$ & $1.4$ & $2.8$ & $7.9$ & $-40$ & $-9.2$ \\
\hline
\end{tabular}
\caption{The critical masses $\mu_{\rm GL}$ for the
Gregory-Laflamme instability and the constant $\gamma$ determining
the behavior of the non-uniform branch for $| \mu - \mu_{\rm
GL}| \ll 1$. \label{tabnonuni}}
\end{center}
\end{table}

In \cite{Sorkin:2004qq} the larger $d$ behavior of $\mu_{\rm GL}$ is also
examined numerically. It is found that the $\mu_{\rm GL}$ versus $d$ curve
for $d \leq 50$ obeys $\mu_{\rm GL} \simeq 16.21 \cdot 0.686^d$ to a good
approximation. That $\mu_{\rm GL}$ behaves exponentially for large $d$
is confirmed by \cite{Kol:2004pn} where
the large $d$ behavior of $\mu_{\rm GL}$ is found analytically
to be $\log \mu_{\rm GL} \simeq d \log \sqrt{ \frac{e}{2\pi} }$.

The qualitative behavior of the non-uniform string branch depends
the sign of $\gamma$. If $\gamma$ is positive, we have that
the branch starts out from $\mu=\mu_{\rm GL}$
with increasing $\mu$ and decreasing $n$.
If $\gamma$ is negative the branch instead starts out from $\mu=\mu_{\rm GL}$
with decreasing $\mu$ and decreasing $n$.
Moreover, one can use the Intersection Rule of \cite{Harmark:2003dg}
(see Section \ref{sec:phase}) to see that
if $\gamma$ is positive (negative) then the entropy of the uniform
string branch for a given mass is higher (lower) than the entropy
of the non-uniform string branch for that mass.
This can also be derived directly from \eqref{nofmu} using
\eqref{neutfirst2}. This gives
\begin{equation}
\frac{\ms_{\rm nu} ( \mu )}{\ms_{\rm u}  ( \mu )}
= 1 - \frac{(d-2)^2}{2(d-1)(d-3)^2} \frac{\gamma}{\mu_{\rm GL}}
(\mu-\mu_{\rm GL})^2 + \CO ( (\mu - \mu_{\rm GL})^3 ) \ ,
\end{equation}
from which we clearly see the significance of the sign of
$\gamma$. Here $\ms_{\rm u} ( \mu )$ ($\ms_{\rm nu} ( \mu )$)
refers to the rescaled entropy of the uniform (non-uniform) black
string branch. We have listed the numerical data for $\gamma$ for
$4 \leq d \leq 14$ in Table \ref{tabnonuni}. From these we see
that for $d \leq 12$ we have that $\gamma$ is positive, while for
$d \geq 13$ we have that $\gamma$ is negative
\cite{Sorkin:2004qq}. Therefore, as discovered in
\cite{Sorkin:2004qq}, we have qualitatively different behavior of
the non-uniform black string branch for small $d$ and large $d$,
i.e. the system exhibits a critical dimension $D=14$.%
\footnote{The first occurrence of a critical dimension in this
system was given in \cite{Kol:2002xz}, where evidence was given
that the merger point between the black hole and the string
depends on a critical dimension $D=10$, such that for $D<10$ there
are local tachyonic modes around the tip of the cone (the
conjectured local geometry close to the thin ``waist'' of the
string) which are absent for $D>10$.}

In six dimensions, i.e. for $d=5$, Wiseman found in
\cite{Wiseman:2002zc} a large body
of numerical data for the non-uniform string branch.
These data are displayed in the $(\mu,n)$ phase diagram
in the right side of Figure \ref{fig1}.
This was originally done in \cite{Harmark:2003dg}.
While the branch starts
in $\mu = \mu_{\rm GL}$  the data ends around $\mu \simeq 2.3 \, \mu_{\rm
GL} \simeq 5.3$.

In the recent paper \cite{Kudoh:2004hs} numerical evidence has
been found that suggest that the non-uniform string branch more or
less ends where the data of \cite{Wiseman:2002zc} end, i.e. around
$\mu \simeq 5.3$ for $d=5$, supporting the considerations of
\cite{Kol:2003ja}. As we discuss more below, this suggests that
the branch has a topology changing transition into the localized
black hole branch.

\subsection*{Localized black hole branch}

On physical grounds we clearly expect that there exists a branch
of neutral black holes in the space-time $\CM^d \times S^1$. One
defining feature of these solutions is that their event horizons
should have topology $S^{d-1}$. We call this branch the {\sl
localized black hole branch} in the following since the $S^{d-1}$
horizon is localized on the $S^1$ of the Kaluza-Klein space.

Neutral black hole solutions in the space-time $\CM^3 \times S^1$
were found and studied in
\cite{Myers:1987rx,Bogojevic:1991hv,Korotkin:1994dw,Frolov:2003kd}.
However, the study of black holes in the space-time $\CM^{d}
\times S^1$ for $d \geq 4$ has only recently begun. A reason for
this is that it is a rather difficult problem to solve since such
black holes are not algebraically special \cite{DeSmet:2002fv}
and, in particular, the solution cannot be found using a Weyl ansatz
since the number of Killing vectors is too small.

\subsubsection*{\underline{Analytical results:}}

Progress towards finding an analytical solution for the localized
black hole was made in \cite{Harmark:2002tr} where, as reviewed
above, the ansatz \eqref{ansatz} was proposed. As stated above,
it was subsequently proven in \cite{Wiseman:2002ti,Harmark:2003eg}
that the localized black hole can be put in this ansatz.

In \cite{Harmark:2003yz} the metric for small black holes, i.e.
for $\mu \ll 1$, was found using the ansatz \eqref{ansatz} of
\cite{Harmark:2002tr} to first order in $\mu$. The first order
metric of the localized black hole branch was also found
analytically in Ref.~\cite{Gorbonos:2004uc}, using a different
method. An important feature of the localized black hole solution
is that $n \rightarrow 0$ for $\mu \rightarrow 0$. This means that
the black hole solution becomes more and more like a
$(d+1)$-dimensional Schwarzschild black hole as the mass goes to
zero.

One of the main results of \cite{Harmark:2003yz} is the first
correction to the relative tension $n$ as function of $\mu$ for
the localized black hole branch, which was found to be%
\footnote{Here $\zeta(p) = \sum_{n=1}^\infty n^{-p}$ is the
Riemann zeta function.}
\begin{equation}
\label{bhslope}
n = \frac{(d-2)\zeta(d-2)}{2(d-1)\Omega_{d-1}} \mu + \CO (\mu^2) \ .
\end{equation}
Using this in \eqref{neutfirst2} one can find the leading correction to the
thermodynamics as
\begin{equation}
\label{cors} \ms_{\rm bh} (\mu) = C_1^{(d)} \mu^{\frac{d-1}{d-2}} \left( 1
+ \frac{\zeta(d-2)}{2(d-2)\Omega_{d-1}} \,  \mu
 + \CO ( \mu^2 )\right) \ ,
\end{equation}
where $C_1^{(d)}$ is defined in \eqref{Cdef}.
This constant of integration is fixed by the physical requirement
that in the limit of vanishing mass we should recover the thermodynamics
of a Schwarzschild black hole in $(d+1)$-dimensional
Minkowski space.
For $d=4$, the second order correction to the metric has recently
been studied \cite{Karasik:2004ds}.

\subsubsection*{\underline{Numerical results:}}

The black hole branch has been studied numerically,
for $d=4$ in \cite{Sorkin:2003ka,Kudoh:2004hs}
and for $d=5$ in \cite{Kudoh:2003ki,Kudoh:2004hs}.
For small $\mu$, the impressively accurate data of
\cite{Kudoh:2004hs} is consistent with the analytical results
of \cite{Harmark:2003yz,Gorbonos:2004uc,Karasik:2004ds}.
We have displayed the results for $d=4,5$ of \cite{Kudoh:2004hs}
in a $(\mu,n)$ phase diagram in Figure \ref{fig1}.

Amazingly, the work of \cite{Kudoh:2004hs} seems to give an answer
to the question: ``Where does the localized black hole branch
end?''. Several scenarios have been suggested, see
\cite{Harmark:2003eg} for a list. The scenario that
\cite{Kudoh:2004hs} points to, is the scenario suggested by Kol
\cite{Kol:2002xz} in which the localized black hole branch meets
with the non-uniform black string branch in a topology changing
transition point. This seems highly evident from the $(\mu,n)$
phase diagram for $d=5$ in Figure \ref{fig1}, and is moreover
supported by examining the geometry of the two branches near the
transition point, and also by examining the thermodynamics
\cite{Kudoh:2004hs}.

Therefore, it seems reasonable to expect that the localized
black hole branch is connected with the non-uniform string
branch in any dimension. This means we can go from the uniform black string
branch to the localized black hole branch through a connected series of
static classical geometries.
The critical behavior near the point in which the localized black hole
and the non-uniform string branch merge has recently been considered in
\cite{Kol:2005vy,Sorkin:2005vz}.

\subsection*{Phase diagrams for $d=4$ and $d=5$}

\begin{figure}[ht]
\centerline{\epsfig{file=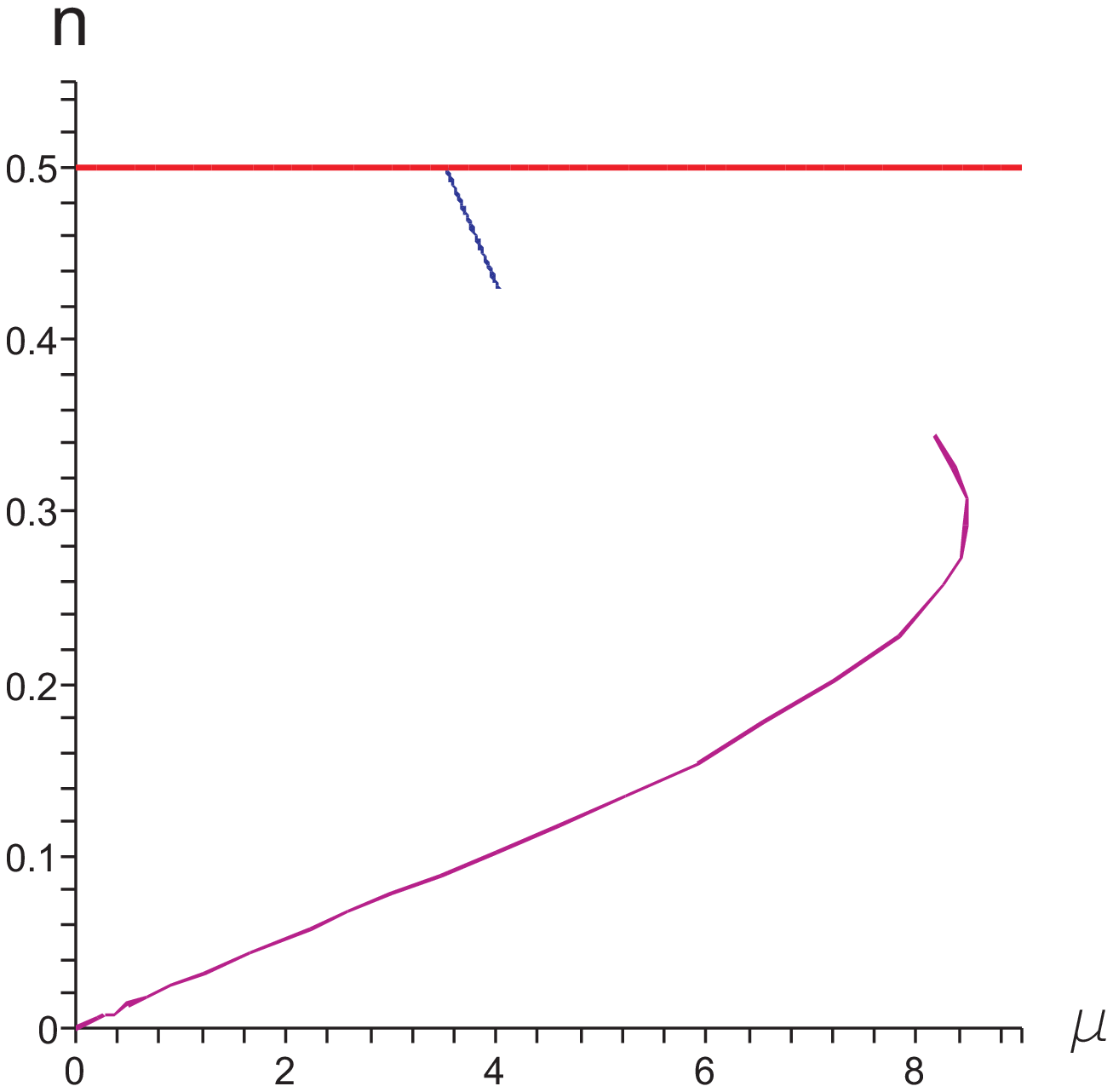,width=7 cm,height=6cm}
\hskip .5cm \epsfig{file=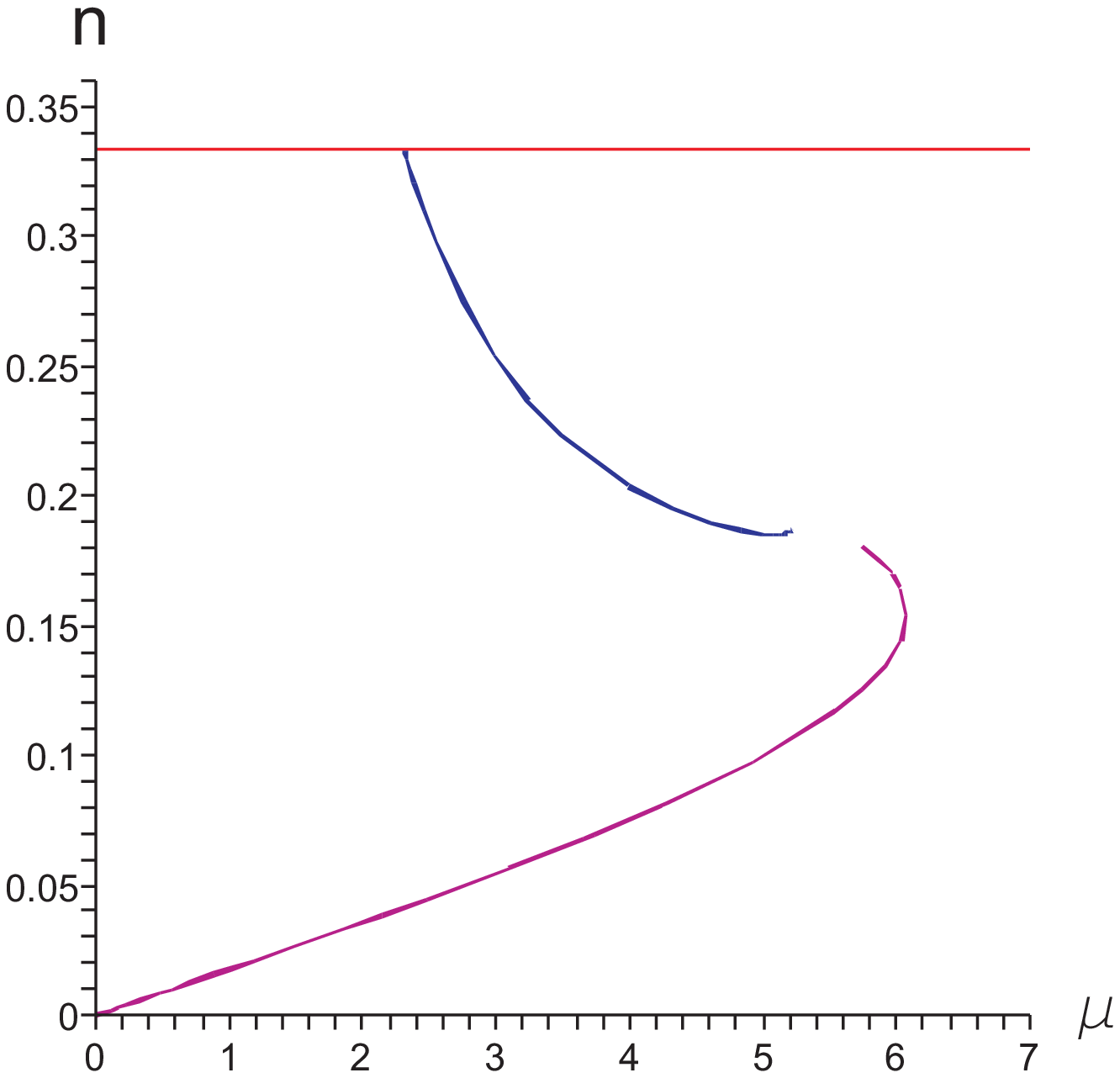,width=7 cm,height=6cm} }
\caption{Black hole and string phases for $d=4$ and $d=5$, drawn in
the $(\mu,n)$ phase diagram. The horizontal (red)
line at $n=1/2$ and 1/3 respectively is the uniform string branch. The (blue)
branch emanating from this at the Gregory-Laflamme mass is the non-uniform
string branch. For $d=4$ only the linear behavior close to the Gregory-Laflamme
mass is known, while for $d=5$ the entire behavior has been obtained numerically
by Wiseman \cite{Wiseman:2002zc}. The (purple) branch starting in the point
$(\mu,n)=(0,0)$ is the black hole branch   which was numerically obtained by
Kudoh and Wiseman \cite{Kudoh:2004hs}. In particular for $d=5$ we observe
the remarkable result that the black hole and non-uniform black string branch
meet.}
\label{fig1}
\end{figure}

We display here in Figure \ref{fig1} the $(\mu,n)$ diagram for $d=4$
and $d=5$, since these are the cases where we have the most knowledge of
the various branches of black holes and strings on cylinders. For $d=4$
we only know the leading linear behavior of the non-uniform branch, which
emanates  at $\mu_{\rm GL} = 3.52$ from the uniform branch which is
given by the horizontal line $n=1/2$. For $d=5$ on the other hand, we have shown
the complete non-uniform
branch, as numerically obtained by Wiseman \cite{Wiseman:2002zc}, which
emanates at $\mu_{\rm GL} = 2.31$ from the uniform branch which has $n=1/3$.
These data were first put in a $(\mu,n)$ diagram in Ref.~\cite{Harmark:2003dg}.
For the black hole branch we have plotted in the figure the recently
obtained numerical data of Kudoh and Wiseman \cite{Kudoh:2004hs}, both
for $d=4$ and 5. As seen from the figure this branch has an approximate linear
behavior for a fairly large range of $\mu$ close to the origin and the
numerically obtained slope agrees very well with the analytic result
\eqref{bhslope}.

\subsection*{Copies of solutions}

In \cite{Harmark:2003eg} it is discussed that one can generate new
solutions by copying solutions on the circle several times,
following an idea of Horowitz \cite{Horowitz:2002dc}. This works
for solutions which vary along the circle direction (i.e. the $z$
direction), so it works both for the black hole branch and the
non-uniform string branch. Let $k$ be a positive integer, then if
we copy a solution $k$ times along the circle we get a new
solution with the following parameters:
\begin{equation}
\label{coptrans} \tilde{\mu} = \frac{\mu}{k^{d-3}} \spa \tilde{n}
= n \spa \tilde{\mt} = k \mt \spa \tilde{\ms} =
\frac{\ms}{k^{d-2}} \ .
\end{equation}
See Ref.~\cite{Harmark:2003eg} for the corresponding expression of
the metric of the copies, as given in the ansatz \eqref{ansatz}.
Using the transformation \eqref{coptrans}, one easily sees that
the non-uniform and the localized black hole branches depicted in
Figure \ref{fig1} are copied infinitely many times in the
$(\mu,n)$ phase diagrams.

\subsection*{The endpoint of the decay of the uniform black
string}

As mentioned above, the uniform black string is
classically unstable for $\mu < \mu_{\rm GL}$. A natural
question to ask is: ``What is the endpoint of the classical
instability?''.

The entropy for a small localized black hole branch is much larger
than that of the than the entropy of a black string of same mass,
i.e. $\ms_{\rm bh} (\mu) \gg \ms_{\rm u} (\mu)$ for $\mu \ll 1$,
as can be easily seen by comparing \eqref{su} and \eqref{cors}.
This suggests that a light uniform string will decay to a
localized black hole. However, one can imagine other possibilities,
for example the uniform black string could decay to another static geometry,
or it could even keep decaying without reaching an endpoint.

\begin{figure}[ht]
\centerline{\epsfig{file=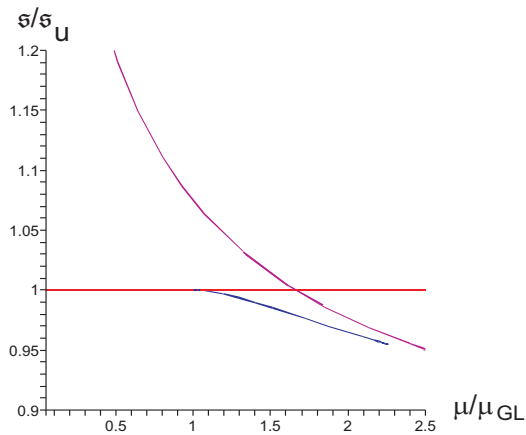,width=7 cm,height=6cm}}
\caption{Entropy $\ms/\ms_{\rm u}$ versus the mass
$\mu/\mu_{GL}$ diagram for the uniform string (red),
non-uniform string (blue) and localized black hole (purple)  branches.}
\label{fig_entr}
\end{figure}

For $d=5$ we can be more precise about this issue.
In Figure \ref{fig_entr} we have displayed the entropy $\ms$ versus the mass
$\mu$ diagram for the localized black hole, uniform string and non-uniform
string branches. We see from this that in six dimensions the
localized black hole always has greater entropy than
the uniform strings, in the mass range where the uniform string is
classically unstable.
Combining this with the fact that the non-uniform string branch
does not exist in this mass range, it suggest that the
unstable uniform black string decays classically to the localized
black hole branch in six dimensions.

The viewpoint that an unstable uniform string decays to a
localized black hole has been challenged in
\cite{Horowitz:2001cz}. Here it is shown that in a classical
evolution an event horizon cannot change topology, i.e. cannot
pinch off, in finite affine parameter (on the event horizon).

However, this does not exclude the possibility that a classically
unstable horizon pinches off in infinite affine parameter. Indeed,
in \cite{Garfinkle:2004em} the numerical study
\cite{Choptuik:2003qd} of the classical decay of a uniform black
string in five dimensions was reexamined, suggesting that the
horizon of the string pinches off in infinite affine parameter.

Interestingly, the classical decay of the uniform string is quite
different for $d \geq 13$. As we reviewed above, the results of
\cite{Sorkin:2004qq} show that for $d \geq 13$ the non-uniform
string has decreasing mass $\mu$ for decreasing $n$, as it
emanates from the uniform string in the Gregory-Laflamme point at
$\mu=\mu_{\rm GL}$. This in addition means that the entropy of the
a non-uniform black string is higher than the entropy of a uniform
string with same mass. Therefore, for $d \geq 13$ we have a
certain range of masses in which the uniform black string is
unstable, and for which we have a non-uniform black string with
higher entropy. This obviously suggests that a uniform black
string in that mass range will decay classically to a non-uniform
black string. Such a decay can then be possible in a finite affine
parameter, according to \cite{Horowitz:2001cz}, since the horizon
topology is fixed in the decay.

Note that the range of masses for which we have a non-uniform
string branch with higher entropy is extended by the fact that we
have copies of the non-uniform string branch. The copies, which
have the quantities given by the transformation rule
\eqref{coptrans}, can easily be seen from the Intersection Rule of
\cite{Harmark:2003dg} (see Section \ref{sec:phase}) to have higher entropy than that of a uniform
string of same mass, since they also have decreasing $\mu$ for
decreasing $n$. Thus, for $d \geq 13$, it is even possible that
there exists a non-uniform black string for any given $\mu <
\mu_{\rm GL}$, with higher entropy than that of the uniform black
string with mass $\mu$. This will occur if the non-uniform string
branch extends to masses lower than $2^{3-d} \mu_{\rm GL}$.
Otherwise, the question of the endpoint of the decay of the
uniform black string will involve a quite complicated pattern of
ranges.

\subsection*{More on large $d$ behavior}

\begin{figure}[ht]
\centerline{\epsfig{file=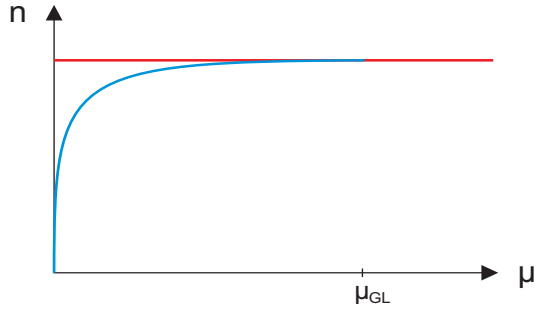,width=7 cm,height=4cm}}
\caption{Sketch of the $(\mu,n)$ phase diagram for large $d$.}
 \label{fig_sketch} \end{figure}

It is interesting to consider the slope of the localized
black hole branch \eqref{bhslope} for large $d$, measured
in units normalized with respect to the Gregory-Laflamme
point $(\mu_{\rm GL},n_{\rm GL}= 1/(d-2))$,
with $\mu_{\rm GL}$ given by the large $d$ formula
of \cite{Kol:2004pn} reviewed above. We get
\begin{equation}
\frac{n}{n_{\rm GL}} \simeq  d^{d/2}
\frac{\mu}{\mu_{\rm GL}} \ ,
\end{equation}
for $d \gg 1$.
We see from this that the slope become infinitely steep
as $d \rightarrow \infty$. This suggests that the curve
describing the localized black hole and non-uniform black string
branches should behave as sketched in Figure \ref{fig_sketch}. Thus,
the large $d$ behavior of the phase structure is expected to be such
that the non-uniform string gets closer and closer to having
$n=n_{\rm GL}= 1/(d-2)$.

\section{Phases with Kaluza-Klein bubbles \label{sec:bub}}

Until now all the solution we have been discussing lie in the range $0 \leq n \leq
1/(d-2)$. But, there are no direct physical restrictions on $n$
preventing it from being in the range $1/(d-2) < n \leq d-2$. The
question is therefore whether there exists solutions in this range
or not. This was answered in \cite{Elvang:2004iz} where it was
shown that pure gravity solutions with Kaluza-Klein bubbles
can realize all values of $n$ in the range $1/(d-2) < n \leq d-2$.

In this section we first review the static Kaluza-Klein bubble
and its place in the $(\mu,n)$ phase diagram. We then discuss the
main properties of bubble-black hole sequences, which are phases of
Kaluza-Klein black holes that involve Kaluza-Klein bubbles. In particular,
we comment on the thermodynamics of these solutions. Subsequently, we present
the five- and six-dimensional phase diagrams as obtained by including
the simplest bubble-black hole sequences. Finally, we comment on
non-uniqueness in the phase diagram.

\subsection*{Static Kaluza-Klein bubble}

Kaluza-Klein bubbles were discovered in \cite{Witten:1982gj} by
Witten. In \cite{Witten:1982gj} it was explained that the
Kaluza-Klein vacuum $\CM^4 \times S^1$ is unstable
semi-classically, at least in absence of fundamental fermions. The
semi-classical instability of $\CM^4 \times S^1$ is in terms of a
spontaneous creation of expanding Kaluza-Klein bubbles, which are
Wick rotated 5D Schwarzschild black hole solutions. The
Kaluza-Klein bubble is essentially a minimal $S^2$ somewhere in
the space-time, i.e. a ``bubble of nothing''. In the expanding
Kaluza-Klein bubble solution the $S^2$ bubble then expands until
all of the space-time is gone. However, apart from these
time-dependent bubble solutions there are also solutions with
static bubbles, as we now first review.

To construct the static Kaluza-Klein bubble in $d+1$ dimensions we
take the Euclidean section of the $d$-dimensional Schwarzschild
black hole and add a trivial time-direction. This gives the metric
\begin{equation}
ds^2 = - dt^2 + \left( 1- \frac{R^{d-3}}{r^{d-3}} \right) dz^2 +
\frac{1}{1- \frac{R^{d-3}}{r^{d-3}}} dr^2 + r^2 \Omega_{d-2}^2 \ .
\end{equation}
We see that there is a minimal $(d-2)$-sphere of radius $R$
located at $r=R$. To avoid a conical singularity we need that $z$
is a periodic coordinate with period
\begin{equation}
\label{Lbub}
L = \frac{4\pi R}{d-3} \ .
\end{equation}
Clearly, the solution asymptotes to $\CM^d \times S^1$ for $r
\rightarrow \infty$. We see that the only free parameter in the
solution is the circumference of the $S^1$.

Now, since the static Kaluza-Klein bubble is a static solution of
pure gravity that asymptotes to $\CM^d \times S^1$ it belongs to
the class of solutions we are interested in. Thus, it is part of
our phase diagram, and using \eqref{gttzz}, \eqref{MT},
\eqref{then}, \eqref{themu} and \eqref{Lbub} we can read off $\mu$
and $n$ as
\begin{equation}
\label{munstat} \mu = \mu_{\rm b} = \Omega_{d-2} \left( \frac{d-3}{4\pi}
\right)^{d-3} \spa n = d-2 \ .
\end{equation}
We see from this that the static Kaluza-Klein bubble is a specific
point in the $(\mu,n)$ phase diagram, which also follows from the
fact that it does not have any free dimensionless parameters.
Notice that $n=d-2$ precisely saturates the upper bound on $n$ in
\eqref{nbound}. In fact, a test particle at infinity will not
experience any force from the bubble (in the Newtonian limit).

The static Kaluza-Klein bubble is known to be classically
unstable. This can be seen from the fact that the static bubble is
the Euclidean section of the Schwarzschild black hole times a
trivial time direction.%
\footnote{Another way to understand the classical instability of
the static Kaluza-Klein bubble is to relate it to the marginal mode
of the uniform black string. See e.g. Ref.~\cite{Sarbach:2004rm}
where the linear stability of
static bubble solutions of Einstein-Maxwell theory was examined. Here
a unique unstable mode was found and shown to be related, by double
analytic continuation, to marginally stable stationary modes of
perturbed charged black strings.}
The Euclidean flat space $\R^3\times S^1$
(hot flat space) is semi-classically unstable to nucleation of
Schwarzschild black holes. This was shown by Gross, Perry, and
Yaffe \cite{Gross:1982cv}, who found that the Euclidean
Lichnerowicz operator $\Delta_\rom{E}$ for the Euclidean section
of the four dimensional Schwarzschild solution with mass $M$ has a
negative eigenvalue: $\Delta_\rom{E} u_{ab} = \lambda u_{ab}$ with
$\lambda=-0.19 (GM)^{-2}$. The Lichnerowicz equation for the
perturbations of the Lorentzian static bubble space-time is
$\Delta_\rom{L} h_{ab}=0$ (in the transverse traceless gauge), so
taking the ansatz $h_{ab} = u_{ab} \, e^{i\Omega t}$, the
Lichnerowicz equation requires $\Omega^2=\lambda$, ie.\
$\Omega=\pm i \sqrt{-\lambda}$.

The classical instability of the static Kaluza-Klein bubble causes
the bubble to either expand or collapse exponentially fast.
For five-dimensional Kaluza-Klein
space-times, there exists initial data \cite{Brill:1991qe} for
massive bubbles that are initially expanding or collapsing
\cite{Corley:1994mc}, and numerical studies \cite{Sarbach:2003dz}
shows that there exist massive expanding bubbles and furthermore
indicates that contracting massive bubbles collapse to a black
hole with an event horizon.

If the bubble is collapsing the endstate is presumably an object
with an event horizon. In this connection, it is noteworthy that
the value of $\mu$ in \eqref{munstat} is smaller than $\mu_{\rm
GL}$, i.e. the Gregory-Laflamme mass, for $4 \leq d \leq 9$ (as
can be seen by comparing $\mu_{\rm b}$ in   \eqref{munstat} to
$\mu_{\rm GL}$ in  Table \ref{tabnonuni}). This means that the
static Kaluza-Klein bubble does not decay to the uniform black
string. It is therefore likely that the Kaluza-Klein bubble in
that case decays to whatever is the endstate of the uniform black
string (see end of Section \ref{sec:bhc}).

\subsection*{Bubble-black hole sequences}

We have now a solution at $n=d-2$, where it should be emphasized that
this solution does not have any event horizon, and hence no entropy or temperature.
But so far in this review we have not mentioned any solutions lying in the
range $1/(d-2) < n < d-2$. However, as shown in \cite{Elvang:2004iz},
the solutions in that range have the property that they contain both
Kaluza-Klein bubbles and black hole event horizons with various topologies.

For $d=4,5$ Emparan and Reall constructed in \cite{Emparan:2001wk}
exact solutions describing a black hole attached to a Kaluza-Klein
bubble using a generalized Weyl ansatz, describing axisymmetric static space-times
in higher-dimensional gravity.%
\footnote{See Ref.~\cite{Harmark:2004rm} for the generalization of
this class to stationary solutions.} For $d=4$ this was
generalized in \cite{Elvang:2002br} to two black holes plus one
bubble or two bubbles plus one black string. There, it was also
argued that the bubble balances the gravitational attraction
between the two black holes, thus keeping the configuration in
static equilibrium.

In \cite{Elvang:2004iz} these solutions were generalized to a
family of exact metrics for configurations with $p$ bubbles and
$q=p,p\pm 1$ black holes in $D=5,6$ dimensions. These are regular
and static solutions of the vacuum Einstein equations, describing
sequences of Kaluza-Klein bubbles and black holes placed
alternately, e.g.~for $(p,q)=(2,3)$ we have the sequence: \beastar
  \rom{black~hole}~-~\rom{bubble}~-~\rom{black~hole}
  ~-~\rom{bubble}~-~\rom{black~hole} \, .
\eeastar
This class of solutions is called \emph{bubble-black
hole sequences} and we refer to particular elements of this class as
$(p,q)$ solutions. This large class of solutions, which was
anticipated in Ref.~\cite{Emparan:2001wk}, thus includes as particular
cases the $(1,1)$, $(1,2)$ and $(2,1)$ solutions obtained and
analyzed in \cite{Emparan:2001wk,Elvang:2002br}.

We refer the reader to \cite{Elvang:2004iz} for the explicit
construction of these bubble-black hole sequences and a comprehensive
analysis of their properties. Here we list a number of essential
features:
\begin{itemize}
\item All values of $n$ in the range $1/(d-2) < n < d-2$ are
realized. \item The mass $\mu$ can become arbitrarily large, and
for $\mu \rightarrow \infty$ we have $n \rightarrow 1/(d-2)$.
\item The solutions contain bubbles and black holes of various
topologies. In the five-dimensional case we find black holes with
spherical $S^3$ and ring $S^2 \times S^1$ topology, depending on
whether the black hole is at the end of the sequence or not.
Similarly, in the six-dimensional case we find black holes with
ring topology $S^3 \times S^1$ and tuboid topology $S^2 \times S^1
\times S^1$, depending on whether the black hole is at the end of
the sequence or not. The bubbles support the $S^1$'s of the
horizons against gravitational collapse. \item The  $(p,q)$
solutions are subject to constraints enforcing regularity, but
this leaves $q$ independent dimensionless parameters allowing for
instance the relative sizes of the black holes to vary.  The
existence of $q$ independent parameters in each $(p,q)$ solution
is the reason for the large degree of non-uniqueness in the
$(\mu,n)$ phase diagram, when considering bubble-black hole
sequences.
\end{itemize}

An interesting feature of the bubble-black hole sequences is that
there exists a  map between five- and six-dimensional
solutions \cite{Elvang:2004iz}. As a consequence there is a
corresponding map between the physical parameters which reads
\begin{equation}
\label{mun5to6} \mu^{(6D)} = \frac{2\pi}{3L^{(5D)}} \mu^{(5D)}
\left( 5 - n^{(5D)} \right) \spa n^{(6D)} =
\frac{5n^{(5D)}-1}{5-n^{(5D)}} \ ,
\end{equation}
\begin{equation}
\label{ts5to6} \mathfrak{t}^{(6D)}_k = \mathfrak{t}^{(5D)}_k \spa
\mathfrak{s}^{(6D)}_k = \frac{4\pi}{L^{(5D)}}
\mathfrak{s}^{(5D)}_k \ ,
\end{equation}
where the superscripts $5D$ and $6D$ label the five- and six-dimensional
quantities respectively. This form of the map assumes a certain normalization of the parameters
of the solution, or equivalently, a choice of units, as further
explained in Ref.~\cite{Elvang:2004iz}.

\subsection*{Thermodynamics}

For static space-times with more than one black hole horizon we
can associate a temperature to each black hole by analytically
continuing the solution to Euclidean space and performing the
proper identifications needed to make the Euclidean solution
regular where the horizon was located in the Lorentzian solution.
The temperatures of the black holes need not be equal, and one can
derive a generalized Smarr formula that involves the temperature
of each black hole. The Euclidean solution is regular everywhere
only when all the temperatures are equal. It is always possible to
choose the $q$ free parameters of the $(p,q)$ solution to give a
one-parameter family of regular equal temperature solutions, which
we denote $(p,q)_{\mathfrak{t}}$.

The equal temperature $(p,q)_{\mathfrak{t}}$ solutions  are of
special interest for two reasons: First, the two solutions,
$(p,q)_{\mathfrak{t}}$ and $(q,p)_{\mathfrak{t}}$, are directly
related by a double Wick rotation which effectively interchanges
the time coordinate and the coordinate parameterizing the
Kaluza-Klein circle. This provides a duality map under which
bubbles and black holes are interchanged, giving rise to the
following explicit map between the physical quantities of the
solutions
\begin{equation}
\label{dW2} \mu' = n  \mt^{d-3} \mu \spa n' = \frac{1}{n} \spa
\mathfrak{t}' = \frac{1}{\mathfrak{t}} \spa \mathfrak{s}' =
\frac{(d-2)n-1}{d-2-n} \mt^{d-1} \mathfrak{s} \ .
\end{equation}
Secondly, for a given family of $(p,q)$ solutions,
the equal temperature solution extremizes the entropy for fixed
mass $\mu$ and fixed size of the Kaluza-Klein circle at infinity.
For all explicit cases considered we find that the entropy is
minimized for equal temperatures.%
\footnote{This is a feature that is particular to black holes,
independently of the presence of bubbles. As an analog, consider
two Schwarzschild black holes very far apart. It is
straightforward to see that for fixed total mass, the entropy of
such a configuration is minimized when the black holes have the
same radius (hence same temperature), while the maximal entropy
configuration is the one where all the mass is located in a single
black hole.}

Furthermore, the entropy of the $(1,1)$ solution is always lower
than the entropy of the uniform black string of the same mass
$\mu$. We expect that all other bubble-black hole sequences
$(p,q)$ have entropy lower than the $(1,1)$ solution, and this
is  confirmed for all explicitly studied examples.
The physical reason to expect that all bubble-black hole sequences
have lower entropy than a uniform string of same mass, is that
some of the mass has gone into the bubble rather than the black
holes, giving a smaller horizon area for the same mass.

\subsection*{Phase diagrams for $d=4$ and $d=5$}

The general $(\mu,n)$ phase diagrams for $d=4,5$ can in principle
be drawn with all possible values $(p,q)$, though the explicit solution
of the constraints becomes increasingly complicated for high $p,q$.
However, the richness of the phase structure and the non-uniqueness in the
$(\mu,n)$ phase diagram, is already illustrated by considering some
particular examples of the general class of solutions, as was
done in \cite{Elvang:2004iz}.
As an illustration, we give here the exact form of the curve
for the (1,1) solution, corresponding to a bubble on a black hole,
\begin{equation}
\label{nmu11} d=4: \quad n_{(1,1)}(\mu )
= \frac{1}{4} \left[ -1 + 3 \sqrt{1 + \frac{8}{\mu^2}} \right] \qquad
; \qquad d=5: \quad
n_{(1,1)} (\mu ) =\frac{1}{3}  + \frac{4}{3\mu} \, .
\end{equation}
in five and six dimensions respectively. These two solutions
are related by the map in \eqref{mun5to6} and one may also check
that these curves are correctly self-dual under the duality
map \eqref{dW2}.

\begin{figure}[ht]
\centerline{\epsfig{file=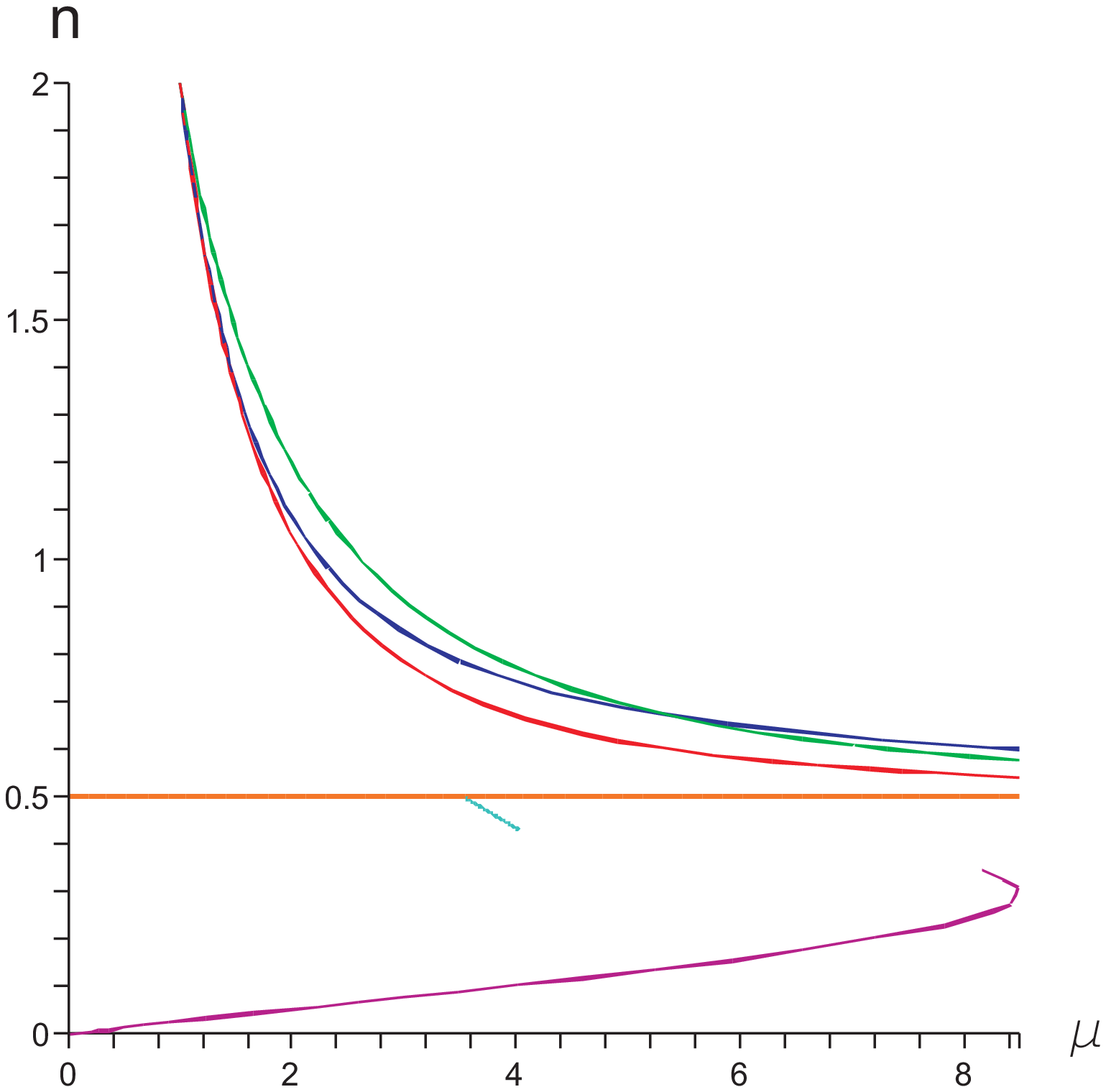,width=7cm,height=7cm}
\hskip .5cm \epsfig{file=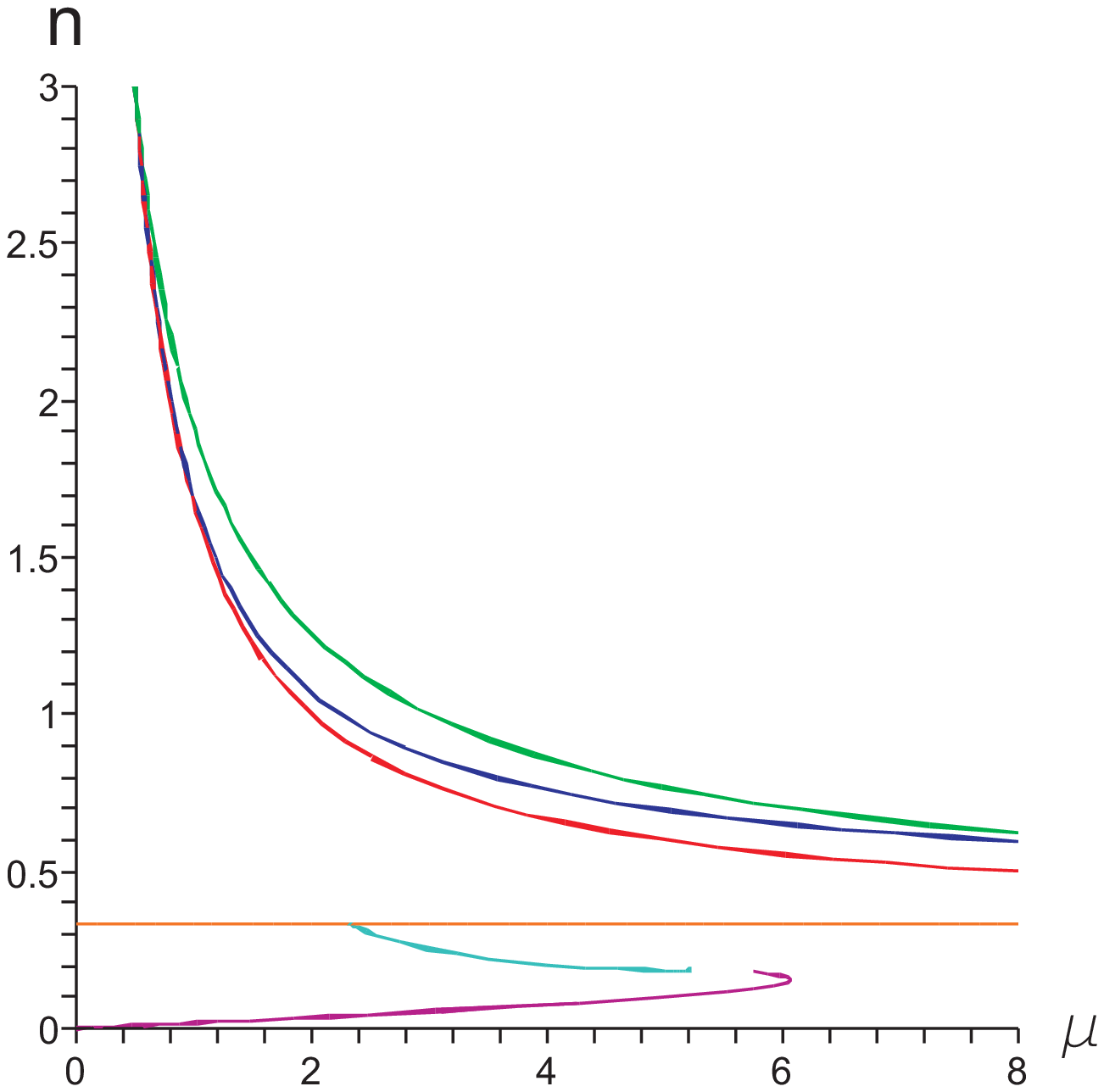,width=7 cm,height=7cm}}
\caption{$(\mu,n)$ phase diagrams for five (left figure) and six
(right figure) dimensions. We have drawn the $(p,q)=(1,1)$,
$(1,2){}_{\mathfrak{t}}$ and $(2,1)$ solutions. These curves lie
in the region $1/2 < n \leq 2$ for the five-dimensional case and
$1/3 < n \leq 3$ for the six-dimensional case. The lowest (red)
curve corresponds to the $(1,1)$ solution. The (blue) curve that
has highest $n$ for high values of $\mu$ is the equal temperature
$(1,2){}_{\mathfrak{t}}$ solution. The (green) curve that has
highest $n$ for small values of $\mu$ is the $(2,1)$ solution. The
entire phase space of the $(1,2)$ configuration is the wedge
bounded by the equal temperature $(1,2){}_{\mathfrak{t}}$ curve
and the $(1,1)$ curve. For completeness we have also included the
uniform (orange) and non-uniform (cyan) black string branch, and
the small black hole branch (magenta) displayed in
Figure \ref{fig1}.} \label{fig2}
\end{figure}

In Figure \ref{fig2} we have drawn for $d=4$ and 5 the curves  in
the $(\mu,n)$ phase diagram for the
$(p,q)=(1,1)$, $(1,2){}_{\mathfrak{t}}$ and $(2,1)$ solutions.
These correspond to the configurations
\begin{equation}
\label{11config}
 \begin{array}{lccc}
 (1,1): \qquad &  \rom{black~hole}&-&\rom{bubble} \\
 D=5 & S^3 &&  D \\
 D=6 &  S^3 \times S^1 &&  D \times S^1
  \end{array}
\end{equation}
\begin{equation}
\label{12config}
 \begin{array}{lccccc}
(1,2) : \qquad  & \rom{black~hole}&-&\rom{bubble} &-& \rom{black~hole} \\
 D=5 & S^3 &&  S^1 \times I   & & S^3 \\
 D=6  & S^3 \times S^1 &&   T^2 \times I && S^3 \times S^1
  \end{array}
\end{equation}
\bea
\label{21config}
   \begin{array}{lccccc}
 (2,1): \qquad & \rom{bubble}&-&\rom{black~ring}&-&\rom{bubble}  \\
 D=5 & D  &&   S^2 \times S^1 && D \\
 D=6 &  D \times S^1 &&   S^2 \times T^2 && D \times S^1
  \end{array}
\eea
where the first/second line in each configuration
corresponds to the topology in five/six dimensions. Here $D$ denotes the
disc and $I$ the line interval.

\subsubsection*{Non-uniqueness in the phase diagram}

We finally remark on non-uniqueness in the $(\mu,n)$
phase diagram for Kaluza-Klein black holes.%
\footnote{Non-uniqueness in higher dimensional pure gravity has
also been found for stationary black hole solutions in
asymptotically flat space-time: Here, there exists for a certain
range of parameters both a rotating black hole with $S^3$ horizon
\cite{Myers:1986un} and rotating black rings with $S^2 \times S^1$
horizons \cite{Emparan:2001wn}.} Clearly for a given mass there is
a high degree of non-uniqueness. The non-uniqueness is not lifted
once we also take into account the relative tension $n$, as there
are explicitly known cases of physically distinct solutions with
the same mass and tension. For example the
$(1,2){}_{\mathfrak{t}}$ solution and the $(2,1)$ solution
intersect each other in the phase diagram. This means that we have
two physically different solutions in the same point of the
$(\mu,n)$ phase diagram.

Moreover, we have in fact a continuously infinite non-uniqueness%
\footnote{Infinite non-uniqueness has also been found in
\cite{Emparan:2004wy} for black rings with dipole charges in
asymptotically flat space.} for certain points in the $(\mu,n)$
phase diagram. This is due to the fact that the $(p,q)$ solution
has $q$ free parameters \cite{Elvang:2004iz}. Hence, for given $p
\geq 2$ and $q \geq 3$, we have $q-2$ free continuous parameters
labelling physically different $(p,q)$ solutions, for certain
points in the $(\mu,n)$ phase diagram.

\section{Discussion}

In this review we  have summarized the current state of knowledge
on the phases of Kaluza-Klein black holes. It is an open question
whether there exist other phases beyond the ones discussed above,
but for five and six dimensions it could be that the picture
presented above is complete.
 There are, however, a number of important issues and
questions regarding the present phase structure, which we briefly list here:
\begin{itemize}
\item At present the complete non-uniform branch is numerically known
only for $d=5$. It would be interesting to compute it in other dimensions.
Similarly, for the black hole branch, the present status is
that only for $d=4$ and 5 do we numerically know the entire branch, and
having the numerical data for other dimensions would be interesting as well.
\item If possible, it would obviously be very interesting to find
an analytic form for the non-uniform and black hole branch, for example
using the ansatz \eqref{ansatz}. A first step in this direction would be
to extend the first order analytical results for these branches to second
order.%
\footnote{For the $d=4$ black hole branch this was done in
Ref.~\cite{Karasik:2004ds}.} \item The classical stability of all
branches (except the uniform one, where we know the Gregory-Laflamme
instability) should be examined.
\item The bubble-black hole
sequences are only known in five and six dimensions. It is
interesting to ask if solutions for bubble-black hole sequences
exist for $D \geq 7$, and if so, whether they can be related to
the solutions of \cite{Elvang:2004iz} by a map similar to the one
mapping five- to six-dimensional solutions. For spaces with more
than one Kaluza-Klein circle, i.e.~asymptotics $\CM^{d}\times T^{k \geq 2}$,
it is easy construct solutions describing regular bubble-black hole sequences
\cite{Elvang:2004iz}.
On the other hand, finding bubble-black hole sequences in $\CM^{d \geq 6}\times S^1$
may be difficult since one cannot use the generalized Weyl ansatz
for such solutions, but we do not see any physical obstructions to
the existence of them. It would also be interesting to see what
type of topologies occur in these higher dimensional bubble-black
hole sequences, should they exist. \item In connection with the
above points, the general question of the $d$-dependence of the
phase diagram is also an important one. For example, for the
non-uniform branch it is known that there is a critical dimension
beyond which the slope changes sign. Likewise one could imagine
new phases appearing (or other phases disappearing) as the number
of dimensions increases.
\item The instability of the uniform black string implies that a
naked singularity may be formed when the horizon of the  black string pinches.
As originally pointed out by Gregory and Laflamme \cite{Gregory:1993vy,Gregory:1994bj},
this would entail a violation of the Cosmic Censorship Hypothesis.
The results of \cite{Wiseman:2002zc,Kudoh:2004hs} suggest that, for
certain dimensions, the localized black hole is the only solution with higher
entropy than that of the uniform black string (for masses where the
black string is unstable). Thus, the endpoint of the instability seems to be
the localized black hole which means that the horizon should pinch off
in the classical evolution. On the other hand, in \cite{Horowitz:2001cz}
it was argued that the horizon cannot pinch off in finite
affine parameter, thus suggesting that it is not possible for the black string
horizon to pinch off.
A way to reconcile these two results is if the horizon pinches off in
infinite affine parameter. Recently, the numerical analysis of
\cite{Choptuik:2003qd,Garfinkle:2004em} indicates that this indeed is the case.
If this is correct  it would be interesting to examine the implications for
the Cosmic Censorship Hypothesis.
\end{itemize}

We conclude with a brief discussion of the relevance
of the study of Kaluza-Klein black holes in the context of
String/M-theory as well as supersymmetric Yang-Mills (SYM) theories and
Little String Theory (LST) (see also the short review \cite{Harmark:2005xx}).
Recently a map%
\footnote{For the class of solutions described by the ansatz \eqref{ansatz},
this map was already discovered in
Ref.~\cite{Harmark:2002tr}.}
was found \cite{Harmark:2004ws}
(see also Refs.~\cite{Bostock:2004mg,Aharony:2004ig} for related
work) from static and neutral Kaluza-Klein black holes to non- and
near-extremal (singly charged) branes of String/M-theory on a transverse circle.
This gives a precise connection between
phases of static and neutral Kaluza-Klein black holes and the
thermodynamic
behavior of the non-gravitational theories dual to near-extremal
branes on a circle. In this way any phase of Kaluza-Klein black holes
has immediate consequences for the phase structure of these
dual non-gravitational theories. In particular,
for the thermodynamics of strongly-coupled SYM theories
 on a circle this has led to
the prediction of a new non-uniform phase as well as new information about
the localized phase \cite{Harmark:2004ws}. For finite temperature two-dimensional SYM
theory on a circle it was shown in \cite{Aharony:2004ig} that a
non-uniform phase at weak coupling indeed appears, with qualitatively
similar features as the one predicted at strong coupling.
As another application, Ref.~\cite{Harmark:2004ws} also presents evidence for
the existence of a new stable phase of $(2,0)$  LST in the canonical ensemble
for temperatures above its Hagedorn temperature.
The application of this map to the bubble-black hole sequences,
by which non and near-extremal bubble-black hole sequences are generated,
is in progress \cite{Harmark:2004bb}.

\section*{Acknowledgments}

We thank Henriette Elvang for collaboration on part of the topics
presented here, and for comments and suggestions to this review.
We thank Hideaki Kudoh and Toby Wiseman for kindly providing their
data on the non-uniform black string and black hole branches in
five and six dimensions and for their help and explanations. We
thank Barak Kol for comments and suggestions to this review. Work
partially supported by the European Community's Human Potential
Programme under contract MRTN-CT-2004-005104 `Constituents,
fundamental forces and symmetries of the universe'.


\providecommand{\href}[2]{#2}\begingroup\raggedright\endgroup


\end{document}